\documentclass[aps,twocolumn, pra, showpacs,superscriptaddress]{revtex4}
\usepackage{graphicx}
\usepackage{dcolumn}
\usepackage{bm}
\usepackage{amsmath}

\begin{document}

\title{Hard-core Bose-Fermi mixture in one-dimensional split traps}
\author{Xiaolong L\"{u}}
\affiliation{Institute of Theoretical Physics, Shanxi University, Taiyuan 030006,
People's Republic of China}
\author{Xiangguo Yin}
\affiliation{Institute of Physics, Chinese Academy of Sciences, Beijing 100086, People's
Republic of China}
\author{Yunbo Zhang}
\email{ybzhang@sxu.edu.cn}
\affiliation{Institute of Theoretical Physics, Shanxi University, Taiyuan 030006,
People's Republic of China}
\date{\today}

\begin{abstract}
We consider a strongly interacting one-dimensional (1D) Bose-Fermi mixture
confined in a hard wall trap or a harmonic oscillator trap with a tunable $\delta$-function
barrier at the trap center. The mixture consists of 1D Bose gas with
repulsive interactions and of 1D noninteracting spin-aligned
Fermi gas, both species interacting through hard-core
interactions. Using a generalized Bose-Fermi mapping, we calculated the
reduced single-particle density matrix and the momentum distribution of the
gas as a function of barrier strength and the parity of particle number. The
secondary peaks in the momentum distribution show remarkable correlation
between particles on the two sides of the split.
\end{abstract}

\pacs{03.75.Mn, 67.85.Pq, 37.10.Gh}
\maketitle

\section{Introduction}

In recent years the strongly interacting ultra-cold atoms system in
quasi-one dimensional(1D) have attracted great interests both in experiments
and theories. With two perpendicular optical lattices to realize the 1D
system and Feshbach resonance or the confinement-induced resonance to tune
the 1D effective interacting strength, Tonks-Girarideau(TG) gas \cite%
{Paredes,Toshiya}, a Bose--Einstein condensate in which the repulsive
interactions between bosonic particles dominate the physics of the system,
has been realized in experiments. For strong attractive interactions between
atoms, the Super Tonks-Girardeau gas \cite{Haller} represents an excited
quantum gas phase in a 1D spatial geometry. Sudden switching from infinitely
strong repulsive to infinitely attractive interactions stabilizes the gas
against collapse and connects the ground state of the Tonks gas to the
excited state of the Super Tonks gas. Meanwhile, the study of a two spin
mixture of 1D strongly interacting fermions gas at finite spin imbalance
demonstrates how ultracold atomic gases in 1D may be used to create
non-trivial new phases of matter experimentally \cite{Moritz05,Mueller}.
Despite intense theoretical and experimental efforts, however, the strongly
interacting Bose-Fermi mixtures remains largely elusive, where interesting
phenomena such as phase separation are predicted to occur \cite%
{Lai,Das,Imambekov,Yin1}.

Most of the theoretical research on Bose-Fermi mixtures so far has been
concentrated on three dimensional systems, and only recently 1D systems
started attracting attention. Theoretical investigations on the quasi-1D
Bose-Fermi mixture so far have focused on the phase diagrams, ground-state
and thermodynamical properties in the scheme of Luttinger liquid theory \cite%
{Cazalilla,Mathey} and Bethe ansatz method \cite{Imambekov,Guan}. For the 1D
ultra-cold atoms system with infinity repulsive interaction between
particles, the Fermi-Bose Mapping (FBM) theorem is used to derive the
wavefunctions and ground state properties of the system. In the simplest
case the mapping function relates the system of impenetrable bosons with
that of non-interacting fermions \cite{Girardeau0} and similar schemes have
been extended to various hard-core systems like spinor Bose gases \cite%
{Deuretzbacher}, spin-1/2 fermions \cite{GuanLiming} and Bose-Fermi mixture
\cite{Girardeau1}.

The starting point of the many-particle solutions of a given
geometry for the TG gas is the exact single-particle eigenstates.
Given the limited examples of exactly solvable single-particle
problems in quantum mechanics, exactly solvable many-particle
problems are also rare even in the TG limit. Here we choose two
typical split traps to load the mixture, i.e., a split hard wall and
a split harmonic oscillator. Investigating the model of the
$\delta$-split potential can be justified in several ways and
related works have been done \cite{Goold,Yin2,Murphy,Lelas}. For
mesoscopic systems, the ground state properties qualitatively differ
for system with a split at the trap center, especially when the
number of particles is odd. On the other hand, $\delta$-split
potential may be viewed as a generic model for double-well
structures or, alternatively, as a good approximation to the problem
of a trap with an impurity at the center. Experimentally this kind
of split is easy to realize by adding an additional laser in the
trap center \cite{Meyrath}. Tunneling of particles through an
energetically forbidden region reveals more and more experimentally
accessible examples of nontrivial quantum phases
\cite{Zollner,Pflanzer,Cai,Hao}.

The paper is organized as follows. After a brief introduction of the
many-body system, in Sec. II we review the Fermi-Bose mapping theorem for
Bose-Fermi mixture and describe the associated single-particle
eigenfunctions and eigenvalues of the $\delta $-split traps. In Sec. III we
calculate the reduced single-particle density matrix and investigate the
influence of the splitting strength on the momentum distribution which is
experimentally accessible. We observe the emergence of bimodal secondary
peaks at the neck of the central peaks for various barrier strengths and
analyze the parity effect of the total number of atoms. Finally, in Sec. IV
we make concluding remarks.

\section{Bosons and Fermions in split traps}

Consider a mixture of $N_{B}$ bosons and $N_{F}$ fermions trapped in
a tight atomic waveguide which restricts the dynamics of the system
into a quasi one-dimensional system in the longitudinal direction.
Assume that the interaction potentials are short-ranged, the
many-particle Hamiltonian at low linear density can be written as
\begin{eqnarray}
H &=&\int dx\left\{ \Psi _{b}^{\dag }\left( -\frac{\hbar ^{2}}{2m_{b}}%
\partial _{x}^{2}+V_{b}(x)\right) \Psi _{b}\right.  \notag \\
&&\left. +\Psi _{f}^{\dag }\left( -\frac{\hbar ^{2}}{2m_{f}}\partial
_{x}^{2}+V_{f}(x)\right) \Psi _{f}\right.  \notag \\
&&\left. +\frac{1}{2}g_{bb}\Psi _{b}^{\dag }\Psi _{b}^{\dag }\Psi _{b}\Psi
_{b}+g_{bf}\Psi _{b}^{\dag }\Psi _{f}^{\dag }\Psi _{f}\Psi _{b}\right\} ,
\label{1}
\end{eqnarray}%
Here, $\Psi _{b}$, $\Psi _{f}$ are the boson and fermion field operators.
The bosonic and fermionic particles are assumed to share the same mass $%
m_{b}=m_{f}=m$, which could be realized by choosing different isotopes of a
given alkali element, e.g. $^{40}K$-$^{39(41)}K$ \cite{Cote} or $^{86(84)}Rb$
- $^{87(85)}Rb$ \cite{Burke}. In our model, the particles experience both
boson-boson ($g_{bb}$) and boson-fermion ($g_{bf}$)\ contact interactions,
while the fermion-fermion interaction is forbidden by the Pauli exclusive
principle ($g_{ff}=0$). We also assume the trapping potential are exactly
the same for both fermions and bosons, $V_{b}(x)=V_{f}(x)=V(x)$.

\subsection{The Fermi-Bose mapping}

The original Fermi-Bose mapping theorem only related strongly interacting
bosons to ideal fermions \cite{Girardeau0}. It was recently found that the
mapping idea can be applied to Fermi Tonks gas and to a series mixture
system as well \cite{Girardeau1}. Concentrating on the situation relevant to
our system, we denote the space coordinates by $X=(x_{1},x_{2},...,x_{N}),$
where $x_{i}$ is the coordinate of a boson when $i\in \left\{
1,...,N_{B}\right\} ,$ or else it labels a fermion provided that $i\in
\left\{ N_{B}+1,...,N_{B}+N_{F}=N\right\} $. Under the hard-core condition $%
g_{bb},g_{bf}\rightarrow \infty $, we may safely drop the two interaction
energies in the Hamiltonian and as a result the many-body wave function of
the system is subject to a constraint that
\begin{equation}
\Phi (X)=0,\text{ \ \ if \ \ \ }x_{i}=x_{j}
\end{equation}%
for $i\neq j$. The first quantized Hamiltonian of our bose-fermi mixture is
then a sum of one-body operators, $H=\sum_{i=1}^{N}h_{i}$, where%
\begin{equation}
h_{i}=-\frac{\hbar ^{2}}{2m}\partial _{x_{i}}^{2}+V\left( x_{i}\right) .
\end{equation}%
Using the Fermi-Bose mapping theorem \cite{Girardeau1}, the total wave
function $\Phi (X)$\ satisfying the constraint is constructed as
\begin{equation}
\Phi (X)=A(X)\Phi _{D}(X),
\end{equation}%
where $\Phi _{D}(X)$ is a Slater determinant of $N$ orthonormal orbitals $%
u_{1}(x),u_{2}(x),\ldots u_{N}(x)$ occupied by $N$ particles%
\begin{equation}
\Phi _{D}(X)=\sum_{P}\varepsilon (P)u_{1}(Px_{1})u_{2}(Px_{2})\ldots
u_{N}(Px_{N}). \label{slater}
\end{equation}%
The sum runs over all $N!$ possible permutations of these variables
including permutations exchanging bosons with fermions, and $\varepsilon
(P)=\pm 1$ for even/odd times of permutation. The symmetry of the system is
repaired by the mapping function $A(X)$%
\begin{equation}
A(X)=\prod_{1\leq j<l\leq
N_{B}}sgn(x_{j}-x_{l})\prod_{j=1}^{N_{B}}%
\prod_{l=N_{B}+1}^{N}sgn(x_{j}-x_{l})  \label{mapfun}
\end{equation}%
so that in general the wavefunction is symmetric under permutations of
bosons and antisymmetric under permutation of fermions. The sign function $%
sgn(x)$ is $+1(-1)$ for $x>0(x<0)$.

\subsection{Eigenstates and eigenvalues of the $\protect\delta $-split traps}

The orthonormal orbitals $u_{n}(x)$ are eigenfuctions of the single-particle
Hamiltonian with eigenvalues $\epsilon _{n}$%
\begin{equation}
\left( -\frac{\hbar ^{2}}{2m}\partial _{x}^{2}+V\left( x\right) \right)
u_{n}(x)=\epsilon _{n}u_{n}(x)
\end{equation}%
with $n=1,2,\ldots N$. In this paper we consider the case that the mixture
is subjected to two kinds of split external trapping potentials, that is,
(A) a hard wall trap which is zero in the region $(-a,a)$ and infinite
outside, and (B) a harmonic oscillator potential with frequency $\omega $,
both with a $\delta $-type barrier located at the origin $x=0$. We denote
the strength of this barrier by a positive parameter $\kappa $. Thus%
\begin{equation}
V\left( x\right) =\left\{
\begin{array}{c}
\kappa \delta \left( x\right) ,\text{ \ }\left\vert x\right\vert <a, \\
\infty ,\text{ \ \ \ \ \ }\left\vert x\right\vert \geq a.%
\end{array}%
\right.
\end{equation}%
for model (A) and%
\begin{equation}
V\left( x\right) =\frac{1}{2}m\omega ^{2}x^{2}+\kappa \delta \left( x\right)
\end{equation}%
for model (B). The single-particle eigenstates of the delta-split hard wall
trap can be found in quantum mechanics textbook, and those of the
delta-split harmonic oscillator have recently been discussed in ref \cite%
{Busch} and we will briefly review the solutions here for completeness.

The eigenfunctions are either symmetric or antisymmetric due to the parity
of both potentials. For model (A), the analytic symmetric eigenfunctions are%
\begin{equation}
u_{n}\left( x\right) =C\left( \cos \left( kx\right) +\frac{m\kappa }{\hbar
^{2}k}\sin (k\left\vert x\right\vert )\right)
\end{equation}%
for $\left\vert x\right\vert <a,$ and $u_{n}\left( x\right) =0$ for $%
\left\vert x\right\vert \geq a$. Here, $C$ is the normalization constant and
$k$ is the wave vector of the particle, determined by $\tan \left( ka\right)
=-\hbar ^{2}k/m\kappa $. The eigenenergies $E=\hbar ^{2}k^{2}/2m$ are given
by the graphical solutions of $k$, that is, the intersection points of the
straight line $-\hbar ^{2}k/m\kappa $ and the tangent function which can be
labeled as $n=1,3,5\ldots $. As $\kappa \rightarrow 0$, the eigenenergies
reduce to those of the ordinary infinite well of width $2a$. As $\kappa
\rightarrow \infty $, the barrier becomes impenetrable, and we have two
isolated infinite square wells of width $a$. By contrast, the antisymmetric
eigenfunctions are zero at the origin, so they never \textquotedblleft
feel\textquotedblright\ the delta function at all. The eigenfunctions and
eigenenergies are exactly the same as those for the infinite square well of
width $2a$%
\begin{equation}
u_{n}\left( x\right) =C\sin \left( kx\right)
\end{equation}%
with $k=n\pi /2a$ and $E_{n}=n^{2}\pi ^{2}\hbar ^{2}/2m(2a)^{2}$ ($%
n=2,4,6,\ldots $). We will use dimensional variables in units of the half
split square potential length $a$, all momenta in units of $p=$ $\hbar /2a,$
and all energies in units of $\pi ^{2}\hbar ^{2}/2ma^{2}$.

For model (B), we immediately know that the antisymmetric eigenfunctions of
the simple harmonic oscillator remain good eigenfunctions for the $\delta $%
-split oscillator, as they vanish at the exact position of the disturbance.
We have
\begin{equation}
u_{n}(x)=N_{n}H_{n}(Q)e^{-Q^{2}/2}
\end{equation}%
where $N_{n}=\left( \sqrt{\pi }2^{n}n!x_{osc}\right) ^{-1/2}$ with $%
n=1,3,5,\ldots $ Here $Q=x/x_{osc}$, $x_{osc}=\sqrt{\hbar /m\omega }$ and $%
H_{n}(Q)$ are the $n$-th order Hermite polynomials. The corresponding
energies are given by the eigenvalues of the odd parity states of the
harmonic oscillator $E_{n}=\left( n+1/2\right) \hbar \omega $.

The even eigenstates of the simple harmonic oscillator, on the other hand,
have an extremum at $x=0$. They are therefore strongly influenced by the
splitting potential and can be found to be \cite{Busch,Goold}%
\begin{equation}
u_{n}(x)=N_{n}e^{-Q^{2}/2}U(-\nu _{n},1/2,Q^{2})
\end{equation}%
where $U(a,b,z)$ are the Kummer's confluent hypergeometric functions \cite%
{Abramowitz} and $\nu _{n}\equiv E_{n}/2\hbar \omega -1/4$ is the
non-integer analog of the principle quantum number $n$ of the
harmonic oscillator. The eigenenergies $E_{n}$ ($n=0,2,4,\ldots $)
are given by the solution of
\begin{equation}
2\Gamma (-\nu _{n}+1/2)+\kappa \Gamma (-\nu _{n})=0
\end{equation}%
Similarly in the following the variables are rescaled in units of the
harmonic-oscillator length $x_{osc}$, the harmonic-oscillator momentum $%
p_{osc}=\hbar /x_{osc}=\sqrt{m\hbar \omega }$, the harmonic-oscillator
energy $\hbar \omega $.

With these single-particle eigenstates, we are now in a position to build
the Slater determinant $\Phi _{D}(X)$ for a system of noninteracting
fermions. The Fermi-Bose mapping theorem then allows us for the calculation
of the exact many particle wave function $\Phi (X)$ from the fermionic
result.

\begin{figure}[tbp]
\includegraphics[width=3.5in]{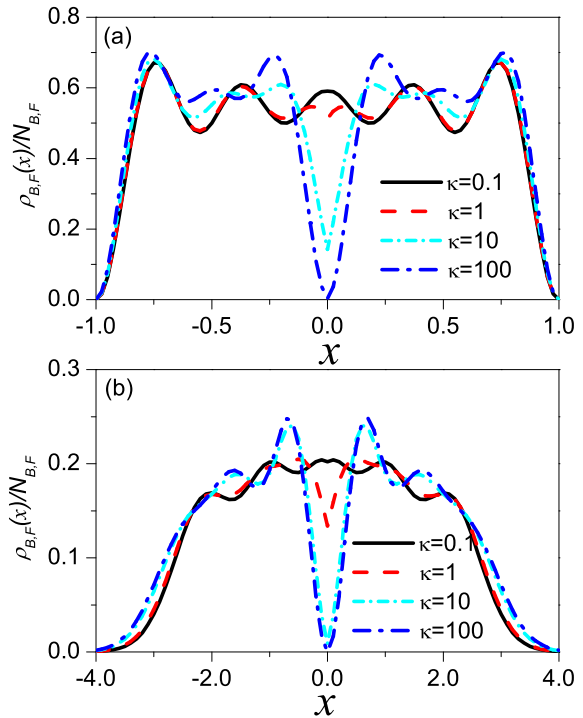}
\caption{(Color online) Total density distributions for odd number
of atoms ($N=N_{B}+N_{F}=5$) at different strengthes of the split
barrier (a) for Model A ($x$ in units of $a$), (b) for Model B ($x$
in units of $\sqrt{\hbar/m \omega}$).} \label{fig1}
\end{figure}

\section{Momentum Distributions of the Bose-Fermi Mixture}

The quantum correlations of the 1D quantum gases can be obtained from the
reduced single-particle density\ (RSPD) matrices defined as%
\begin{eqnarray}
\rho _{B}(x,y) &=&N_{B}\int dX^{\prime }\Phi ^{\ast }(x,X^{\prime })\Phi
^{\ast }(y,X^{\prime })  \notag \\
\rho _{F}(x,y) &=&N_{F}\int dX^{\prime \prime }\Phi ^{\ast }(X^{\prime
\prime },x)\Phi (X^{\prime \prime },y)
\label{rhobf}
\end{eqnarray}%
for bosons and fermions, respectively. Here $X^{\prime
}=(x_{2},...,x_{N})$ and $X^{\prime \prime }=(x_{1},...,x_{N-1})$.
In the TG interaction limit, however, the diagonal elements are
nothing but the single-particle density profiles $\rho
_{B}(x,x)=\rho _{B}(x)$ or $\rho _{F}(x,x)=\rho _{F}(x)$ which are
exactly the same up to normalization factors due to the fact that
the square of the mapping function $A(X)$ in (\ref{mapfun}) is
unity. They both are proportional to the spatial density $\rho
_{TG}(x)$ of a trapped TG gas made of $N$ bosons. In Fig 1 we
illustrate the normalized density
profiles%
\begin{equation}
\rho _{B}(x)/N_{B}=\rho _{F}(x)/N_{F}=\rho _{TG}(x)/N
\end{equation}%
for split hard wall trap (Model A) and split harmonic oscillator
(Model B), respectively. Here we are interested in the effect of the
$\delta $-barrier in the center and results for four values of the
barrier strength $\kappa =0.1,1,10,100$ are presented. It has been
shown that for even number of particles the barrier separates the
particles into two half wells and increasing the barrier strength
leads to the emergence of a quadrant separation of which the
interference is negligible \cite{Yin1,Murphy}. Thus only results for
odd number of particles are given here and we found that the density
profile relies on the total number of particles, instead of the
number of bosons and fermions. For negligibly small split barrier,
the density profiles present exactly the same number of density
maxima as the total number of atoms. The corresponding density
profile in the large-$N$ limit is flat in the square well and is a
semicircle of radius $R=\sqrt{2N}$ in the harmonic trap. Increasing
the barrier strength would gradually cut the density profiles into
two symmetric parts and the number of peaks do not match the number
of particles any more. This is a clear signature of the coherence
inherent in the system, which becomes more pronounced in the
momentum distribution. The delta-splitting seems more efficient in
suppressing the density in a harmonic oscillator. This can be
understood by noticing that the energy unit adopted in hard wall is
$\pi^2/2$ times larger than that in harmonic oscillator.

The Fermi-Bose mapping theorem gives identical density distributions of a
sample for bosons and fermions in the TG limit. However, the momentum
distribution can still be used for distinction, which can be calculated from
the reduced single-particle density matrix%
\begin{equation}
n_{B,F}(p)=\frac{1}{2\pi }\iint \rho _{B,F}(x,y)e^{-ip(x-y)}dxdy
\end{equation}%
which is normalized to $N$. The momentum distribution of the bosonic
component $n_{B}(p)$ is equivalent to that of $N$ TG bosons placed
within the split traps. This again comes from the fact that the
symmetry of both boson-boson and boson-fermion permutations has been
repaired in the mapping function and has already been addressed in
Ref. \cite{Goold,Yin1,Lelas}. The physical reason for this identical
momentum distribution originates from the equal-weighted
superposition of the orthonormal orbitals in the construction of the
many-body wavefunction (\ref{slater}). Thus we now turn our
attention to the behavior of the fermionic momentum distribution in
the mixture $n_{F}(p)$ as a function of the split barrier strengths.
The determination of the fermionic momentum distribution proves very
complex even in the homogeneous mixture and the extension to the
harmonically trapped mixture has been done in Ref. \cite{Fang}. The
calculation is more involved because the expression for the density
matrix could not be reduced to a simpler analytical formula, and we
have resorted to numerical computations. Following a more convenient
scheme, we use the
alternative expression for momentum distributuion%
\begin{equation}
n_{F}(p)=N\int dX^{\prime \prime }\left\vert \tilde{\Phi}(X^{\prime \prime
},p)\right\vert ^{2}
\end{equation}%
Here $\tilde{\Phi}(X^{\prime \prime },p)$ is the Fourier transformation of $%
\Phi (X^{\prime \prime },x_{N})$ with respect to the last fermionic variable
$x_{N}$%
\begin{equation}
\tilde{\Phi}(X^{\prime \prime },p)=\frac{1}{\sqrt{2\pi }}\int_{-\infty
}^{+\infty }dx_{N}\Phi (X^{\prime \prime },x_{N})e^{-ipx_{N}}.
\end{equation}%
In a recent paper, Lelas et al. studied the RSPDM, momentum
distribution, natural orbitals and their occupancies for a
Bose-Fermi mixture in an alternative form of a harmonic potential
with a Gaussian-type barrier in the center, however, for a rather
strong fixed barrier, i.e. the barrier is at least several times
larger than the energy of the $N$-th single-particle state of the
potential. We present here the results for various barrier strengths
and analyze the dependence of the distribution profiles on the
parity of the total number of particles. Furthermore our results
provide a direct comparison of the momentum distribution for
different type of trapping potentials (hard-wall and harmonic
oscillator).

\begin{figure}[tbp]
\includegraphics[width=3.5in]{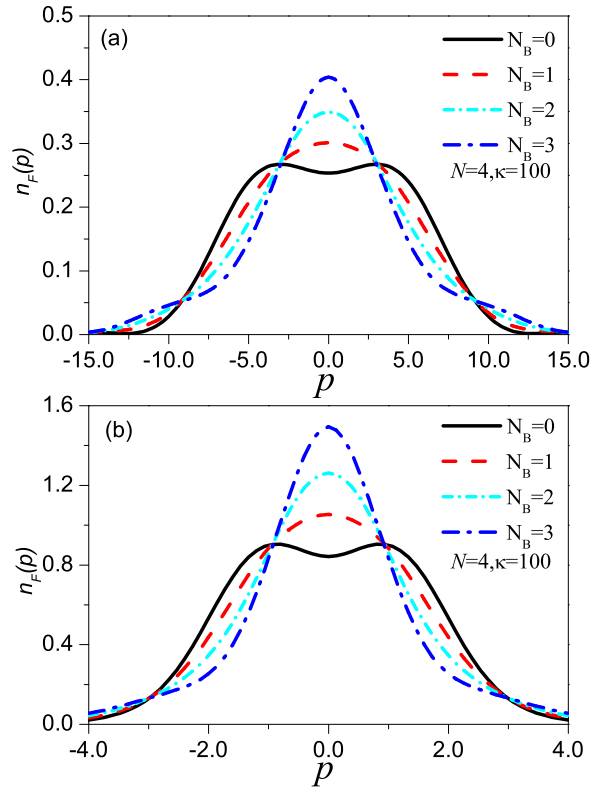}
\caption{(Color online) Momentum distributions for even number of
atoms ($ N=N_B+N_F=4$) at strong split barrier $\protect\kappa =100$
(a) for Model A ($p$ in units of $\hbar/2a$), (b) for Model B ($p$
in units of $\sqrt{m \hbar \omega}$).} \label{fig2}
\end{figure}

Figure 2 shows our results for a mixture with a fixed even total number of $%
N=4$ particles when the number of bosons is increased from $N_{B}=0$ up to $%
3 $. A mixture with $N_{B}=N$ or $N-1$ makes no difference because
one can not imprint any statistical property onto the only left
atom. Both distributions for split hard wall and split harmonic
oscillator take smooth bell-shaped profiles, with explicit
difference being the characteristic momentum up to which the
fermionic momentum distribution is significantly different from
zero. The signature of fermionization becomes more evident for less
bosons. We also note that the split barrier in the center hardly
affect the momentum distribution of even number of atoms, except a
little broadening of the momentum distribution and lowering of its
peak.

\begin{figure}[tbp]
\includegraphics[width=3.5in]{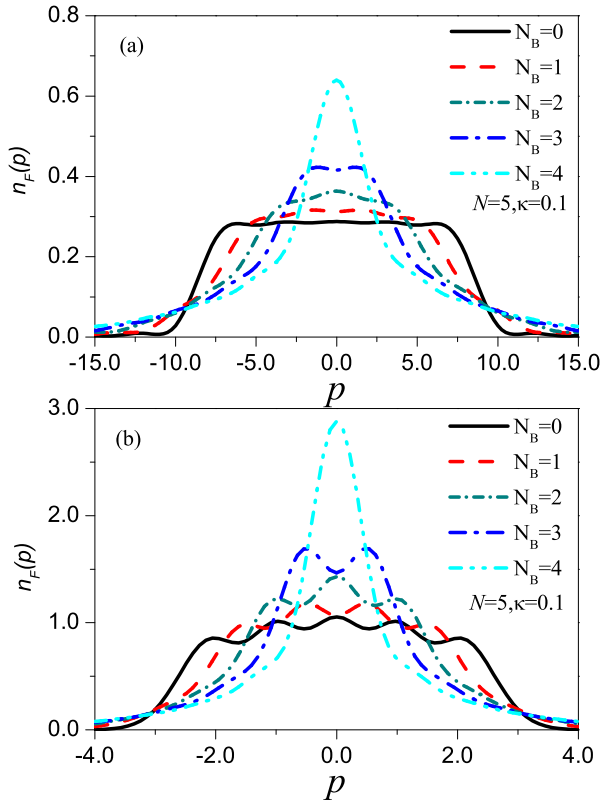}
\caption{(Color online) Momentum distributions for odd number of atoms ($%
N=N_{B}+N_{F}=5$) at weak split barrier $\protect\kappa =0.1$ (a)
for Model A ($p$ in units of $\hbar/2a$), (b) for Model B ($p$ in
units of $\sqrt{m \hbar \omega}$).} \label{fig3}
\end{figure}

Things become much more different for odd number of particles. We
first illustrate the situation for a weak barrier. Figure 3 shows
our results for a BF mixture with a fixed total number of $N=5$
particles when the number of bosons is increased from $N_{B}=0$ up
to $4$ (or, equivalently, $5$). We observe $N_{F}$ \textquotedblleft
fermionic\textquotedblright\ oscillations developed in the momentum
distribution, yet the oscillations in harmonic oscillator are more
prominent. Fermionic oscillations are suppressed for more bosons and
the tails of the distribution become more easily seen. Finally, when
the atoms in the mixture are all bosons, we see typical
Tonks-Girardeau-type momentum distribution.

A strong split barrier greatly modifies the momentum distribution of
odd number of atoms. Quite similar to the bosonic case, one can see
the emergence of bimodal secondary peaks at the neck of the central
peaks in Figure 4, which stems from the interference of the
particles in two almost separate wells. The secondary peaks remain
prominent for all combination of bosons and fermions, moving inward
when the number of bosons is increased. Again we observe that the
characteristic momentum is notably different for the square well and
for the harmonic confinement. For the square well it is of the order
of $k\sim\pi N /2$, while for the harmonic trap it goes as
$k\sim\sqrt{2 N}$. The large-$N$ asymptotic in absence of bosons in
both cases is step function.

\begin{figure}[tbp]
\includegraphics[width=3.5in]{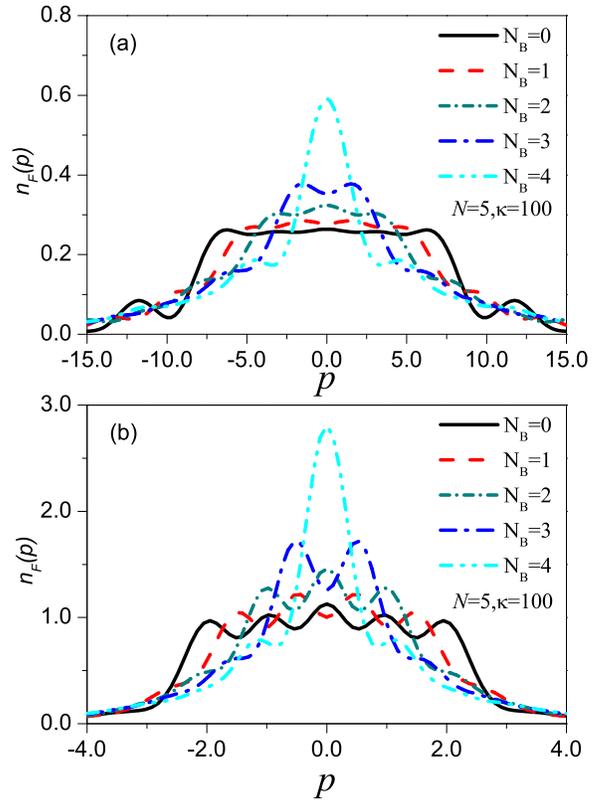}
\caption{(Color online) Momentum distributions for odd number of atoms ($%
N=N_B+N_F=5$) at strong split barrier $\protect\kappa =100$ (a) for
Model A ($p$ in units of $\hbar/2a$), (b) for Model B ($p$ in units
of $\sqrt{m \hbar \omega}$).} \label{fig4}
\end{figure}

\section{Conclusion}

In conclusion, we have shown the ground state properties of a
Bose-Fermi mixture in split potential wells. We observe that the
most significant difference between the total even and odd numbers
of particles occurs in the emergence of bimodal secondary peaks in
the momentum distributions. Three critical features can be seen (1)
the momentum distribution depends on the ratio of the bosons and
fermions in the mixture, as well as the total number of the atoms;
(2) the presence of a strong split barrier at the trap center
enhance greatly the correlation of the atoms on the two separated
wells; (3) the dependence of characteristic momentum on the number
of atoms relies on the shape of trapping potential.

\begin{acknowledgments}
This work is supported by NSF of China under Grant No. 10774095, NSF
of Shanxi Province under grant No. 2009011002, National Basic
Research Program of China (973 Program) under Grant Nos.
2006CB921102 and 2010CB923103, and Program for New Century Excellent
Talents in University (NCET).
\end{acknowledgments}


\begin{thebibliography}{99}


\bibitem{Paredes} B. Paredes, A. Widera, V. Murg, O. Mandel, S. F\"{o}lling,
I. Cirac, G. V. Shlyapnikov, T. W. H\"{a}nsch, and I. Bloch, Nature \textbf{%
429}, 277 (2004).

\bibitem{Toshiya} T. Kinoshita, T. Wenger, and D. S. Weiss, Science \textbf{%
305}, 1125 (2004).

\bibitem{Haller} E. Haller, M. Gustavsson, M. J. Mark, J. G. Danzl, G. Hart,
G. Pupillo, and H.-C. N\"{a}gerl, Science \textbf{325}, 1224 (2009).

\bibitem{Moritz05} H. Moritz, T. St\"{o}ferle, K. G\"{u}nter, M. K\"{o}hl,
and T. Esslinger, Phys. Rev. Lett. \textbf{94}, 210401 (2005).

\bibitem{Mueller} Y. Liao, A. S. C. Rittner, T. Paprotta, W. Li, G. B.
Partridge, R. G. Hulet, S. K. Baur, and E. J. Mueller, arXiv:0912.0092.

\bibitem{Lai} C. K. Lai, C. N. Yang, Phys. Rev A \textbf{3}, 393 (1971); C.
K. Lai, J. Math. Phys. \textbf{15}, 954 (1974).

\bibitem{Das} K. K. Das, Phys. Rev. Lett. \textbf{90}, 170403(2003).

\bibitem{Yin1} X. Yin, S. Chen, and Y. Zhang, Phys. Rev. A \textbf{79},
053604 (2009).

\bibitem{Imambekov} A. Imambekov, E. Demler, Ann. Phys. (N.Y.) \textbf{321},
2390 (2006); Phys. Rev. A \textbf{73}, 021602(R) (2006).

\bibitem{Cazalilla} M. A. Cazalilla, A. F. Ho, Phys. Rev. Lett. \textbf{91},
150403 (2003).

\bibitem{Mathey} L. Mathey, D. W. Wang, W. Hofstetter, M. D. Lukin, and E.
Demler, Phys. Rev. Lett. \textbf{93}, 120404 (2004).

\bibitem{Guan} M. T. Batchelor, M. Bortz, X.-W. Guan, and N. Oelkers, Phys.
Rev. A \textbf{72}, 061603(R) (2005); X.-W. Guan, M. T. Batchelor, and J.-Y.
Lee, ibid. \textbf{78}, 023621 (2008).

\bibitem{Girardeau0} M. Girardeau, J. Math. Phys. (N.Y.) \textbf{1}, 516
(1960).

\bibitem{Deuretzbacher} F. Deuretzbacher, K. Fredenhagen, D. Becker, K.
Bongs, K. Sengstock, and D. Pfannkuche, Phys. Rev. Lett. \textbf{100},
160405 (2008).

\bibitem{GuanLiming} L. Guan, S. Chen, Y. Wang, and Z.-Q. Ma, Phys. Rev.
Lett. \textbf{102}, 160402 (2009).

\bibitem{Girardeau1} M. D. Girardeau and A. Minguzzi, Phys. Rev. Lett.
\textbf{99}, 230402 (2007).

\bibitem{Yin2} X. Yin, Y. Hao, S. Chen, and Y. Zhang, Phys. Rev. A \textbf{78},
013604 (2008).

\bibitem{Goold} J. Goold and Th. Busch, Phys. Rev. A \textbf{77}, 063601
(2008).

\bibitem{Lelas} K. Lelas, D. Juki\'{c} and H. Buljan, Phys. Rev. A \textbf{80%
}, 053617 (2009).

\bibitem{Murphy} D. S. Murphy, J. F. McCann, J. Goold, and T. Busch, Phys.
Rev. A \textbf{76}, 053616 (2007).

\bibitem{Meyrath} T. P. Meyrath, F. Schreck, J. L. Hanssen, C.-S. Chuu, and M.
G. Raizen, Phys. Rev. A \textbf{71}, 041604(R) (2005).

\bibitem{Cai} X. Cai, L. Guan, S. Chen, Y. Hao, and Y. Wang, arxiv:0909.5523.

\bibitem{Pflanzer} A. C. Pflanzer, S. Z\"{o}llner, P. Schmelcher, arxiv:0911.5142.

\bibitem{Zollner} S. Z\"{o}llner, H.-D. Meyer, and P. Schmelcher, Phys. Rev. Lett.
\textbf{100}, 040401 (2008); Phys. Rev. A \textbf{78}, 013629 (2008); \textbf{74}, 053612 (2006).

\bibitem{Hao}Y. Hao, Y. Zhang, J.-Q. Liang, and S. Chen, Phys. Rev. A \textbf{73}, 063617 (2006).

\bibitem{Cote} R. C\^{o}t\'{e}, A. Dalgarno, H. Wang and W. C. Stwalley,
Phys. Rev. A \textbf{57}, R4118 (1998).

\bibitem{Burke} J. P. Burke, J. L. Bohn, Phys. Rev. A \textbf{59}, 1303
(1999); S. G. Crane, X. Zhao, W. Taylor and D. J. Vieira, Phys. Rev. A
\textbf{62}, 011402(R) (2000).

\bibitem{Busch} Th. Busch, B. G. Englert, K. Rzazewski, and M. Wilkens,
Found. Phys. \textbf{28}, 549 (1998).

\bibitem{Abramowitz} M. Abramowitz and I. A. Stegun, eds., Handbook of
Mathematical Functions (Dover, New York, 1972).

\bibitem{Fang} B. Fang, P. Vignolo, C. Miniatura and A. Minguzzi, Phys. Rev.
A \textbf{79}, 023623 (2009).



\end{thebibliography}
\end{document}